\documentclass[twocolumn]{biophys-empty}

\usepackage[utf8]{inputenc}
\usepackage{graphicx}

\usepackage{amsmath}
\usepackage{color}
\usepackage{graphicx}
\usepackage{epstopdf}
\usepackage{mathtools}
\usepackage{texdraw}
\usepackage{tikz}
\usepackage{transparent}
\usepackage{textcomp}
\usepackage{gensymb}
\usepackage{graphicx}
\usepackage[separate-uncertainty=true]{siunitx}
\usepackage{placeins}
\usepackage[hidelinks]{hyperref}

\usepackage{lipsum}

\DeclareMathOperator{\sgn}{sgn}
\renewcommand{\d}[1]{\mathrm{d}#1}
\newcommand{\dd}[2]{\frac{\mathrm{d}#1}{\mathrm{d}#2}}

\newcommand{\pdd}[2]{\frac{\partial #1}{\partial #2}}

\renewcommand{\div}[1]{\pmb{\nabla} \cdot #1}

\newcommand{\A}{\pmb{A}}
\newcommand{\B}{\pmb{B}}
\renewcommand{\v}{\pmb{v}}
\newcommand{\w}{\pmb{w}}

\newcommand{\Tw}{T\!w}

\renewcommand{\Im}{I_\mathrm{m}}

\DeclareSIUnit{\molar}{M}
\DeclareSIUnit{\angstrom}{\mbox{\normalfont\r{A}}}

\graphicspath{{Figures/}}

\title{Combined force-torque spectroscopy of proteins\\by means of multiscale molecular simulation}
\runningtitle{Spectroscopy via multiscale simulation}

\author[1,*]{Thijs W. G. van der Heijden}
\author[2]{Daniel J. Read}
\author[2]{Oliver G. Harlen}
\author[1,3]{Paul van der Schoot}
\author[4,5]{Sarah A. Harris}
\author[1,6]{Cornelis Storm}

\runningauthor{van der Heijden, Read, Harlen, van der Schoot, Harris, Storm}

\affil[1]{Theory of Polymers and Soft Matter, Eindhoven University of Technology, P.O. Box 513, 5600 MB Eindhoven, The Netherlands}
\affil[2]{School of Mathematics, University of Leeds, Leeds LS2 9JT, United Kingdom}
\affil[3]{Instituut voor Theoretische Fysica, Universiteit Utrecht, Princetonplein 5, 3584 CC Utrecht, The Netherlands}
\affil[4]{School of Physics and Astronomy, University of Leeds, Leeds LS2 9JT, United Kingdom}
\affil[5]{Astbury Centre for Structural Molecular Biology, University of Leeds, Leeds LS2 9JT, United Kingdom}
\affil[6]{Institute for Complex Molecular Systems, Eindhoven University of Technology, P.O. Box 513, 5600 MB Eindhoven, The Netherlands}

\corrauthor[*]{t.w.g.van.der.heijden@tue.nl}

\papertype{Article}

\begin{document}

\newlength{\figwidth}
\setlength{\figwidth}{0.5\linewidth}
\setlength{\figwidth}{\linewidth}

\begin{frontmatter}

\begin{abstract}
Assessing the structural properties of large proteins is important to gain an understanding of their function in, \textit{e.g.}, biological systems or biomedical applications. We propose a method to examine the mechanical properties of proteins subject to applied forces by means of multiscale simulation. We consider both stretching and torsional forces, which can be applied independently of each other. We apply torsional forces to a coarse-grained continuum model of the antibody protein immunoglobulin~G (IgG) using Fluctuating Finite Element Analysis and identify the area of strongest deformation. This region is essential to the torsional properties of the molecule as a whole, as it represents the softest, most deformable domain. We subject this part of the molecule to torques and stretching forces on an atomistic level, using molecular dynamics simulations, in order to investigate its torsional properties. We calculate the torsional resistance as a function of the rotation of the domain, while subjecting it to various stretching forces. We learn how these obtained torsion profiles evolve with increasing stretching force and show that they exhibit torsion stiffening, which is in qualitative agreement with experimental findings. We argue that combining the torsion profiles for various stretching forces effectively results in a combined force-torque spectroscopy analysis, which may serve as a mechanical signature for the examined molecule.
\end{abstract}

\begin{sigstatement}
{In this work, we propose a multiscale numerical approach to assess the mechanical properties of macromolecules such as proteins. We perform a combined force-torque spectroscopy analysis on the mechanically most relevant domain to compute the response signature of the spatial structure of the macromolecule. This information may lead to a better understanding of molecular structure and function in biological context and may be used towards diagnostic and sensing applications in the biomedical field.}
\end{sigstatement}

\end{frontmatter}

\section*{Introduction}
Proteins fulfil numerous different roles in organisms, such as providing rigidity, transporting cargo through cells or catalysing reactions. The functioning of a protein arises from the folding of its intrinsic structure, a linear chain of amino acids, into a higher-order hierarchical structure. Its final folded structure consists of a certain shape with one or more active sites, which facilitate the protein's function~\cite{Bourne2003}. One type of protein in particular, the antibody or \textit{immunoglobulin}, plays an important role in the mammalian immune system, by specifically binding to foreign structures in the body. The fact that its binding to a particular molecule is very specific makes the antibody protein an excellent candidate to be employed for analyte detection in a so-called immunoassay~\cite{John2013}. In an immunoassay, the analyte is targeted by an antibody molecule equipped with, \textit{e.g.}, a fluorescent or radioactive label. This label can in turn be detected using conventional detection methods.

In order to make immunoassays a viable method for point-of-care diagnostics in medical applications, however, not only the analysis, but also all of the sample preparation, transportation and mixing steps should be included in a `lab-on-a-chip' device. One proposed method for the integration is making use of magnetic particles within the device~\cite{VanReenen2014, Moerland2019}. Not only can these particles serve as labels for the analytes, but upon actuation with magnetic fields they can be actively manipulated and be used to, \textit{e.g.}, mix fluids within the sensor. In addition, the magnetic particles may act as magnetic tweezers in order to exert forces on the molecules~\cite{VanReenen2013, Gutierrez-Mejia2015, Moerland2019}. This experimental method is used to study the structural properties and unfolding of molecules such as proteins, DNA and RNA~\cite{Smith1992, Schemmel1999, Gosse2002, Neuman2008, Long2013}, and may be employed side-by-side with atomic force microscopy (AFM)~\cite{Rief1997, Oesterhelt1999, Rief2002, Puchner2009, Scholl2014} and optical tweezers~\cite{Nishizaka1995, Kellermayer1997, Tskhovrebova1997, Wang1997, Bennink2001, Wen2007} experiments to assess the mechanical properties of molecules.

In recent years, \citet{VanReenen2013} investigated the torsional resistance of immunoglobulin protein complexes using magnetic tweezers experiments. They were able to distinguish between different torsion profiles for different proteins, a quality that may eventually be employed to, \textit{e.g.}, identify the nature of the binding (to be either specific or non-specific) in immunoassays. To gain more insight into the relation between the intrinsic structure of a molecule and its mechanical properties, we propose a multiscale numerical approach to analyse the mechanical properties of proteins, in our case an immunoglobulin molecule, when subjected to externally applied forces. Since investigating such large molecules ($\sim10^5\ \SI{}{\dalton}$) as a whole on an atomistic level is too costly in a computational context, we examine the molecule on a coarse-grained mesoscopic level using Fluctuating Finite Element Analysis (FFEA)~\cite{Oliver2013, Richardson2014, Hanson2015}. FFEA considers the overall shape of the molecule and regards it as a continuum material internally. This allows for a fast evaluation of the molecule subject to thermal and/or external forces, at the expense of the loss of information on the internal structure. Using FFEA, we can identify the area in which immunoglobulin deforms the most during torsion, which is presumably an essential region for the molecule's torsional properties: it constitutes a weak link in the rigidity of the molecule. Using molecular dynamics simulations, we perform a combined force-torque spectroscopy analysis on this domain on a microscopic, full-atom level: we investigate the torsional resistance of the molecule as a function of the rotation, while a stretching force is exerted on the structure. We extract the resulting torsion profile of the molecule and learn how it develops as the stretching force increases.

The remainder of this manuscript is organised as follows. In the Methods section, we describe the subject molecule of this work -- immunoglobulin~G (IgG) -- in more detail. We briefly describe the Fluctuating Finite Element Analysis (FFEA) and molecular dynamics (MD) methods that we use in our simulations. In the Results and Discussion section, we present the results from the FFEA simulations, which we use to define the domain of interest. We discuss our results from the atomistic MD simulations of this relevant domain, in which we subjected it to external forces and torques. In the Conclusions section, we summarise our findings and draw our conclusions.

\section*{Methods}
\subsection*{Force-torque spectroscopy on Immunoglobulin~G}
\label{sec:IgG}
\begin{figure}[htp!]
\centering
\includegraphics[width = \figwidth]{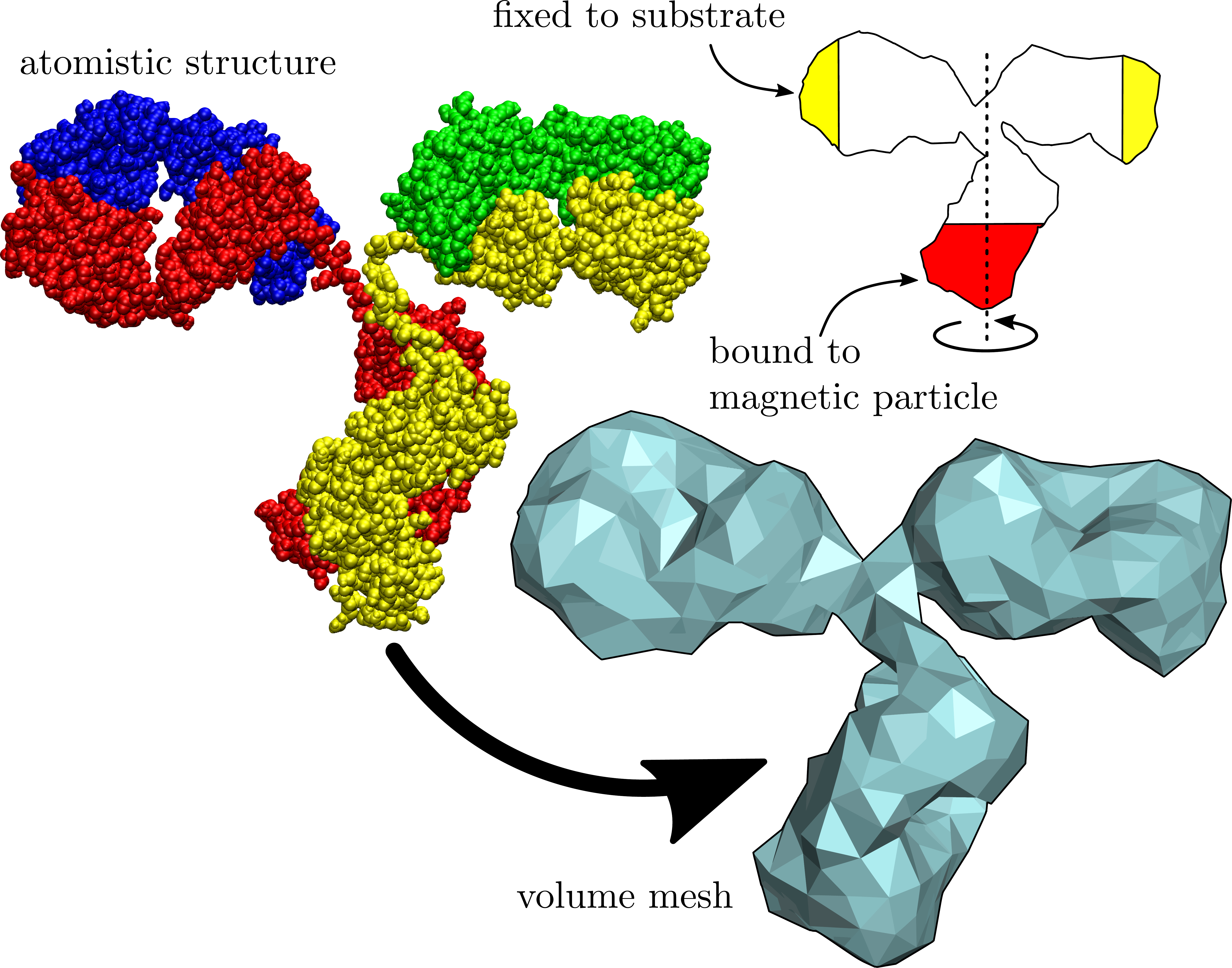}
\caption{The overall shape of the atomistic structure (top left) of IgG is converted into a volume mesh (bottom right). Schematic (top right): the torsional force is exerted on the molecule at the bottom (red shaded area), while the protein is fixed at the ends of the top branches (yellow shaded areas), to mimic the fixation to a substrate. The torsion axis and rotation direction are indicated.}
\label{fig:IgG_discretise}
\end{figure}

Our subject molecule is an antibody protein, immunoglobulin~G (IgG), of which an atomistic structure was found by X-ray diffraction (mouse IgG, Protein Data Bank:~1IGT)~\cite{Harris1997}, see Fig.~\ref{fig:IgG_discretise} (top left). The protein consists of two structurally identical heavy chains (red and yellow) and two identical light chains (blue and green), with a total mass of $\sim\SI{150}{\kilo\dalton}$. The four chains combined form three bulky domains (the three branches of the typical ``Y''-shape), connected by a thin linker; a feature shared by all isotypes of immunoglobulin~\cite{Janeway2001}.

The IgG molecule was the subject of experimental work by~\citet{VanReenen2013}, who investigated the torsional properties of a protein complex, formed by either two IgG molecules or an IgG and a protein~G molecule. They sandwiched the complex between a glass substrate and a magnetic particle and found that the different protein complexes respond differently to exerted torques and that they stiffen for increasing torsion angles: they exhibit a torsion stiffening behaviour. We note that in practice, in addition to the torsional forces, a stretching force may be exerted on a protein complex by applying a second magnetic field that pulls the magnetic particle away from the substrate. However, stretching forces are already inherently present in the experimental situation due to the gravitational forces that work on the magnetic particle and the forces that arise from the direct interaction between the magnetic particle and the substrate to which the protein complex is bound.

Considering the addition of such stretching forces in the torsion experiments allows for a second, independent axis to exert forces on the molecules on. Arguably, both the conventional stretching and torsional rigidities of a molecule depend in fact simultaneously on both the amount of stretching and the amount of torsion. This creates a direct coupling between the stretching and the torsion in such a combined force-torque spectroscopy analysis, and may result in a more complex mechanical ``signature'' of a certain structure. We aim to numerically investigate the mechanical properties of a single IgG molecule, and to calculate the torsion profile of part of IgG under the influence of stretching forces. To that end, we first identify our region of interest on a mesoscopic level using Fluctuating Finite Element Analysis (FFEA). We briefly discuss the method below.

\subsection*{Mesoscopic simulation using Fluctuating~Finite~Element~Analysis (FFEA)}
\label{sec:ffea}
The FFEA method treats large molecules such as proteins as a continuum material, in order to simulate their behaviour on a mesoscopic scale~\cite{Oliver2013, Solernou2018}. The underlying principle is that the overall shape of such a molecule determines its function, and that the intrinsic structure of the molecule is of less importance. We presume that the dynamics of such a material is described by the Cauchy momentum equation,
\begin{equation}
\label{eq:cauchy}
\pdd{\pmb{u}}{t}+\left(\pmb{u}\cdot\pmb{\nabla}\right)\pmb{u}=\frac{1}{\rho}\div{\pmb{\sigma}},
\end{equation}
where $\pmb{u}$ denotes the velocity at any point in the material, $\rho$ is the mass density of the material and $\pmb{\sigma}$ is the total stress exerted by the continuum material. We employ the Kelvin-Voigt material model, which allows us to write the total stress $\pmb{\sigma}$ as the sum of elastic, viscous and thermal stress terms,
\begin{equation}
\label{eq:stresses}
\pmb{\sigma}=\pmb{\sigma}_\mathrm{e} + \pmb{\sigma}_\mathrm{v} + \pmb{\sigma}_\mathrm{t}.
\end{equation}
The elastic stress $\pmb{\sigma}_\mathrm{e}$ is based on a Mooney-Rivlin hyperelastic model for the stress~\cite{Shontz2012, Gent1996}, which is described in more detail in the Appendix. For the expressions for the viscous and thermal stresses we refer the reader to Ref.~\cite{Oliver2013}, since these are of secondary importance in this study. The protein's material properties are parametrised using continuum material parameters, such as the mass density $\rho$, the bulk modulus $K$ and the shear modulus $G$. To directly measure experimentally the values for these parameters is not straightforward, although estimations can be made by considering experimental data on the density of proteins (roughly $\SI{1.5}{\gram/\centi\metre^{3}}$)~\cite{Fischer2004, Quillin2000}, their internal viscosity ($\sim \SI{1}{\milli \pascal \second}$)~\cite{Oliver2013a} and their elastic modulus (order $10^{7}-10^{8}\ \SI{}{\pascal}$)~\cite{Oliver2013a, Voss2015}. We note, however, that the location of the most deformed area is arguably more sensitive to the geometry of the molecule than to the exact values for the material properties, since we exert external forces on the molecule and parametrise the structure homogeneously.

We create the volume mesh for the molecule by considering the atomistic model, converting it to an electron density map and meshing its surface~\cite{Solernou2018}, as illustrated in Fig.~\ref{fig:IgG_discretise}. We coarsen the surface until the shortest edge is $\SI{7}{\angstrom}$, while maintaining the volume~\cite{Oliver2013a}, and create the volume mesh using the Netgen software package. We investigate the IgG molecule in numerical torsion simulations by exerting an external torque on the bottom branch of the molecule, see Fig.~\ref{fig:IgG_discretise}. In order to mimic the fixation to a substrate, we immobilise the ends of the top two branches. We note that we do not exert stretching forces on the molecule in these FFEA simulations. In order to avoid strong sudden deformations of the molecule, we slowly increase the total torque on the molecule up to various constant values $\tau$. We exert the torque by adding an additional torsion force to all the mesh nodes within the red shaded area in Fig.~\ref{fig:IgG_discretise}, where the magnitude depends on the distance to the torsion axis. For definiteness, we disregard the thermal stresses in the material, and rather focus on the viscoelastic response of the structure to the external torque. To that end, we set the thermal stresses in our simulations to zero. 

By studying the internal stresses in the molecule during torsion using FFEA, we find the domain of interest for this protein. We isolate the domain and analyse it using molecular dynamics simulations. We briefly describe the simulations below.

\subsection*{Atomistic simulation using molecular~dynamics (MD)}
\label{sec:md}
We perform molecular dynamics (MD) simulations on the relevant domain of the molecule using the GROMACS software package~\cite{Abraham2015}. The simulation is performed in an implicit water solution, with a \SI{100}{\milli \molar} monovalent salt concentration, using a Langevin thermostat with friction constant $\gamma=\SI{5}{\per\pico\second}$, in order to ensure a strongly damped dynamics. This is prudent in order to suppress the influence of the high rotation rate (as discussed below) and to minimise inertial effects in our analysis of the mechanical response. We employ the Amber ff99SB-ILDN forcefield for the parametrisation of the interactions in the atomistic representation of the protein~\cite{Lindorff-Larsen2010}. 

\begin{figure}[htp!]
\centering
\includegraphics[width=\figwidth]{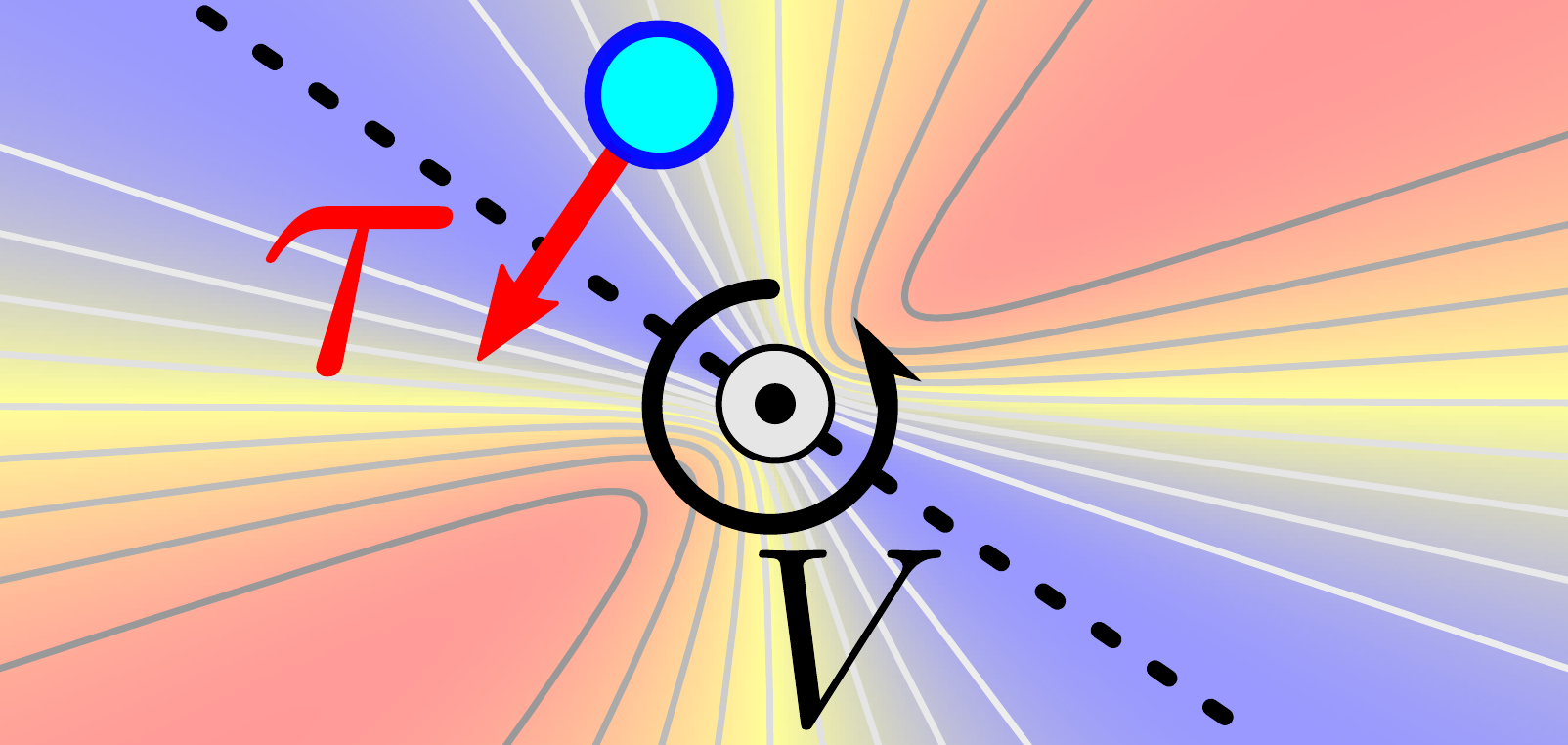}
\caption{Schematic of the potential well $V$, that rotates in a counter-clockwise direction with a rate of $\SI{100}{\degree \per \nano \second}$. Blue indicates low potential energy, red indicates high potential energy. The potential minimum is indicated by the dashed line. The forced residue, indicated by the blue circle, is pulled towards the bottom of the well, resulting in the torque $\tau$. The central axis (the gray dashed line in Fig.~\ref{fig:IgG_linker}) is indicated by $\odot$.}
\label{fig:schematic_potential_well}
\end{figure}

By simultaneously exerting a stretching force $f$ and a torque $\tau$ on the structure we are able to explore its mechanical properties in this combined force-torque spectroscopy analysis. We exert various stretching forces $f$ on the molecule, ranging from $0$ to $\SI{3200}{\kilo \joule \per \mole \per \nano \metre}$ (approximately $\SI{5.3}{\nano \newton}$). We exert the torque $\tau$ on the molecule by rotating a harmonic potential well $V$ around the central axis of the molecule, which the forced residues of the structure are pulled into (GROMACS: $V^{\text{rm2}}$)~\cite{GROMACS_manual}, see Fig.~\ref{fig:schematic_potential_well}.
The potential minimum is indicated by the dashed line, and the blue and red colours represent low and high potential energy, respectively. The rotation rate is $\SI{100}{\degree \per \nano \second}$ and the magnitude of the torque depends on the positions of the residues as well as the spring constant $k$ associated with the potential well. $k$ ranges from $0$ to $\SI{3000}{\kilo \joule \per \mole \per \nano \metre \squared}$ ($\sim \SI{5}{\newton \per \metre}$) in our simulations, resulting in torques $\tau$ up to approximately $\SI{2000}{\kilo \joule \per \mole}$ ($\sim \SI{3300}{\pico \newton \nano \metre}$). For reference, in the experiments by \citet{VanReenen2013}, the maximum exerted torques are of the order of $\SI{4000}{\pico \newton \nano \metre}$.

In the next section, we present our results on the mesoscopic FFEA simulations in order to identify the domain of interest in the molecule. We isolate this area and perform full-atom MD simulations on it, subjecting it to external forces and torques. We analyse the torque as a function of the torsion of the molecule and the exerted stretching force, resulting in a combined force-torque spectroscopy analysis of the domain.

\section*{Results and Discussion}
\subsection*{Mesoscopic FFEA simulations}
\begin{figure}[htp!]
\centering
\large
\includegraphics[width=\figwidth]{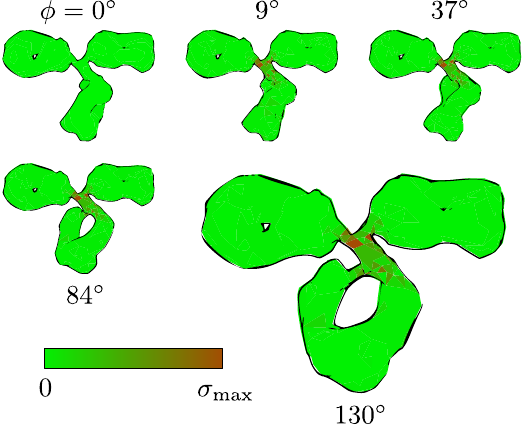}
\caption{Snapshots of a typical FFEA torsion simulation of the IgG molecule for various torsion angles $\phi$, for a total torque $\tau = \SI{299}{\pico \newton \nano \metre}$. The surface elements shaded in red indicate the areas in which the total stress is relatively high. The intensity of the red shade is scaled to the maximum stress magnitude $\sigma_\mathrm{max}$ in each snapshot. The areas of high stress indicate strong deformations.}
\label{fig:IgG_torsion}
\end{figure}

We perform mesoscopic simulations of the IgG molecule subject to torsional forces using FFEA. In Fig.~\ref{fig:IgG_torsion} we show snapshots of a typical torsion simulation for various values of the torsion angle $\phi$. We measure the stress at each position in the molecule and shade the stressed regions in red. The intensity of the red colour indicates the amount of stress, which is normalised to the maximum stress in each snapshot.

We find that for increasing torsion angles $\phi$, the stress in the linker area between the three bulky domains strongly increases. The high stress indicates a strong deformation in this area, which hints at this region being critical for the torsional resistance of the molecule as a whole: it constitutes the softest area in the structure. We note that in this particular case this is not entirely surprising, considering that it is a relatively thin linker in the otherwise bulky geometry of the molecule. However, we argue that in general this need not be the case. The use of the FFEA continuum model to assess the magnitude of the stress as a function of position within a molecule enables us to select which region is necessary to consider for more detailed analysis using atomistic simulations, which account for the internal structure of the protein~\cite{Note}.

We further investigate the linker domain and its mechanical properties on a microscopic level: we isolate the linker region and subject it to torques and stretching forces in molecular dynamics simulations in order to extract its torsion profile for various stretching forces. We discuss the results below.

\subsection*{Atomistic MD simulations}
\begin{figure}[htp!]
\centering
\includegraphics[width=\figwidth]{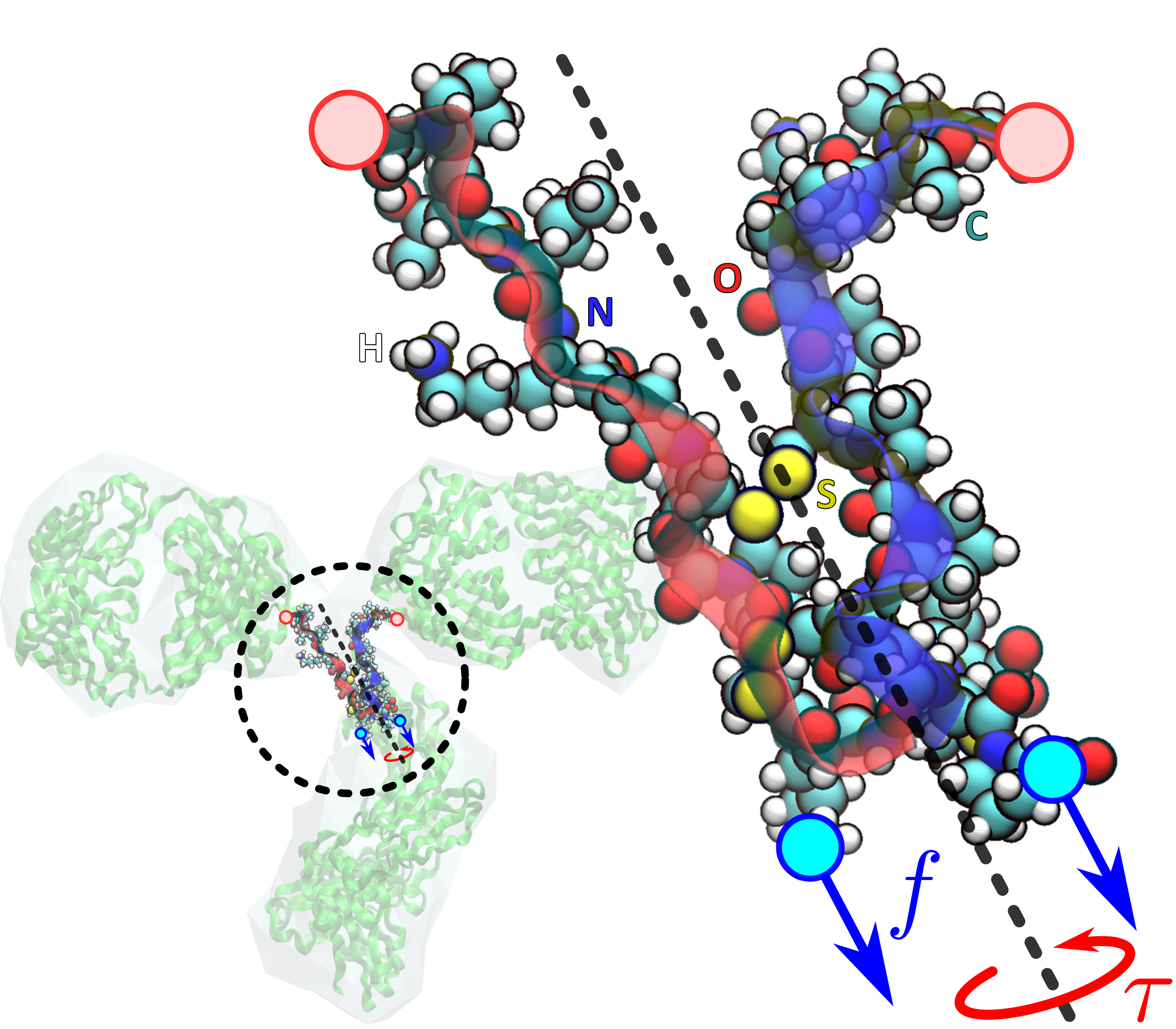}
\caption{The linker region, consisting of two identical protein chains (indicated by the red and blue ribbons), in a full-atom representation. The central axis of the molecule is indicated (gray dashed line), as well as the direction of the torque $\tau$ and the pulling forces $f$. The residues at the top (red circles) are fixed to their position, the torque and force are exerted on the residues at the bottom (blue circles).}
\label{fig:IgG_linker}
\end{figure}

The flexible linker region of the IgG molecule consists of two identical peptide chains with 13 residues each, interconnected by three disulfide bonds, see Fig.~\ref{fig:IgG_linker}. Note that only two of the disulfide bonds (in yellow) are clearly visible.

We first perform an energy minimisation and a NVT simulation, while keeping the ends of the chains immobilised, to equilibrate the molecule. Subsequently, we perform a molecular dynamics simulation while subjecting the molecule to a stretching force $f$ along its central axis, in the absence of a torque. We fix the top residues (indicated in red) in place, to mimic the fact that in reality they are connected to the rather bulky top two branches of the molecule. For the initial stretching, we allow the bottom residues (indicated in blue) to only move strictly in the direction of the stretching force, as this serves as an equilibration for the length of the chains. Then, the bottom residues are released and a torque $\tau$ is exerted on them along the molecule's central axis, which causes the structure to rotate.

Directly calculating the rotation angle $\phi$ of the bottom residues around the torsion axis is only sensible if the molecule remains reasonably stretched during the torsion. Due to the double-chain nature of the structure, however, exerting a strong torque on the molecule not only results in a rotation around the torsion axis, but it also causes additional coiling of the central axis of the molecule. This coiling is not accounted for if we directly measure $\phi$. In order to capture all of the torsion in the structure, we consider the twist $\Tw$ of the backbones of the chains in the linker region as a measure for the amount of rotation contained in the molecule. The twist is a quantity of a mathematical ribbon, independent of an external reference axis, that describes the winding of the ribbon around itself with respect to the ribbon axis~\cite{Au2008}. We note that the value for $\Tw$ becomes negative if we move along the ribbon in the opposite direction. We construct the ribbon by considering the pairs of corresponding atom between the backbones of the two identical chains, from the lowest disulfide bond, up to the top residues. We define the ribbon axis as the average positions of each pair of atoms and the ribbon boundary consists of one of the two peptide chains. The protocol to calculate the twist, as well as the notion that using the writhe as an alternative measure of the torsion of the molecule turns out to not be viable, are discussed in detail in the Appendix.

\begin{figure}[htp!]
\centering
\includegraphics[width=\figwidth]{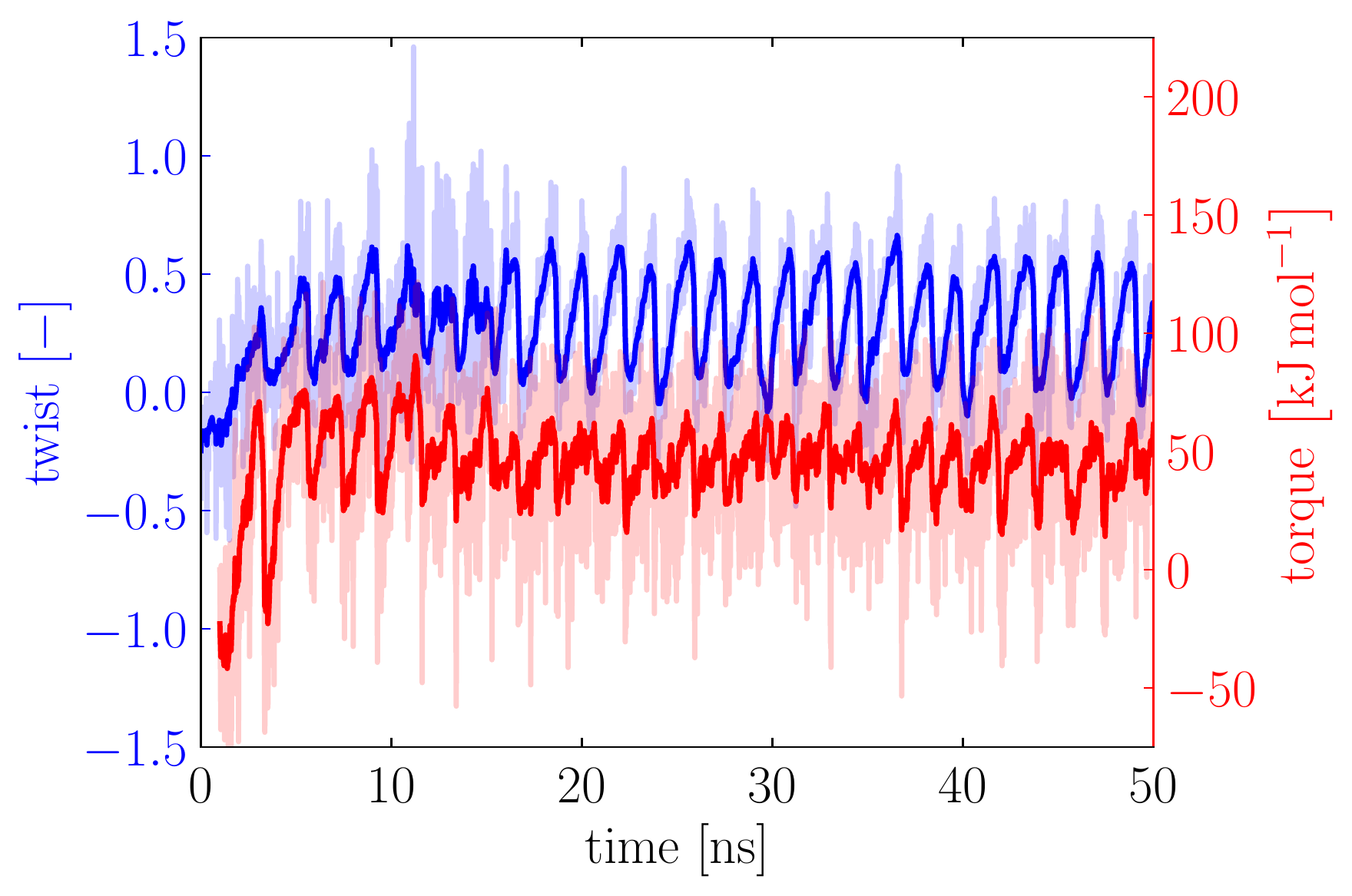}
\caption{The twist $\Tw$ (blue) and applied torque $\tau$ (red) as a function of time, for a stretching force $f=\SI{800}{\kilo\joule\per\mole\per\nano\metre}$ and a spring constant $k=\SI{50}{\kilo\joule\per\mole\per\nano\metre\squared}$. The light colour represents the original data, the darker colour shows a weighted running average of the data over time, calculated using Eq.~\eqref{eq:running_avg}.}
\label{fig:twist_torque_vs_time}
\end{figure}

In our torsion simulations, we calculate the twist $\Tw$ of the structure for each snapshot of the simulation and track the exerted torque $\tau$. Figure~\ref{fig:twist_torque_vs_time} shows the data from a typical simulation, for which we arbitrarily set the stretching force $f=\SI{800}{\kilo\joule\per\mole\per\nano\metre}$ and the spring constant $k=\SI{50}{\kilo\joule\per\mole\per\nano\metre\squared}$, which results in a regularly oscillating behaviour.
We show the data for the twist $\Tw$ in blue, and the torque $\tau$ in red. The light colour represents the original data, the darker colour shows a weighted running average of the data over time, in order to highlight the trends. The weighted running average $\bar{x}_i$ at a point in time $i$ is calculated by taking $2N$ data points surrounding $i$, and weighing them by the inverse distance to the point,
\begin{equation}
\label{eq:running_avg}
\displaystyle \bar{x}_i = \frac{\sum_{j = -N}^{N} x_{i + j} / (|j| + 1)}{\sum_{j = -N}^{N} 1 / (|j| + 1)},
\end{equation}
where $x_i$ is the original data at the point in time $i$. For the trends in Fig.~\ref{fig:twist_torque_vs_time}, we set $N=30$.

We see that the data for both the twist and the torque strongly fluctuate as a result of the Brownian motion of the molecule, but that overall the twist and torque show coherent behaviour. At very short times ($t < \SI{100}{\pico \second}$), we see a fairly constant value for $\Tw$, and no torque $\tau$ is exerted. This corresponds to the length equilibration step, where we stretch the molecule using a force $f$ without exerting a torque. Subsequently, we release the bottom residues and rotate the potential well $V$. As a result, we see an increase of both $\Tw$ and $\tau$ in time. After a while, the molecule reaches a state of oscillatory motion, as both $\Tw$ and $\tau$ reach a maximum value, after which the molecule partially relaxes to a less strained state. The process repeats for each (half-)cycle of the rotating potential well: as long as the rotated residues remain near the potential minimum, hardly any torque is exerted. Upon further rotation, however, the torsional resistance increases and a torque is exerted on the residues in order to enforce the rotation. As the torsional resistance of the molecule then becomes too strong for the potential well to overcome, the residues move out of the well and the structure is allowed to partially relax until the rotating potential catches up to it again. Since the potential well is symmetric with respect to the central axis (as is schematically shown in Fig.~\ref{fig:schematic_potential_well}), this results in a period for the oscillatory motion of half a rotation of the well, corresponding to $\SI{1.8}{\nano \second}$ for a rate of $\SI{100}{\degree \per \nano \second}$.

We repeat the torsion simulations for various values of the spring constant $k$ for the rotating potential well $V$. We track the twist $\Tw$ and the torque $\tau$ during the simulations and combine the results into a torsion profile for a given stretching force $f$: we show the exerted torque $\tau$ as a function of the twist $\Tw$ of the molecule.
\begin{figure}[htp!]
\centering
\includegraphics[width=\figwidth]{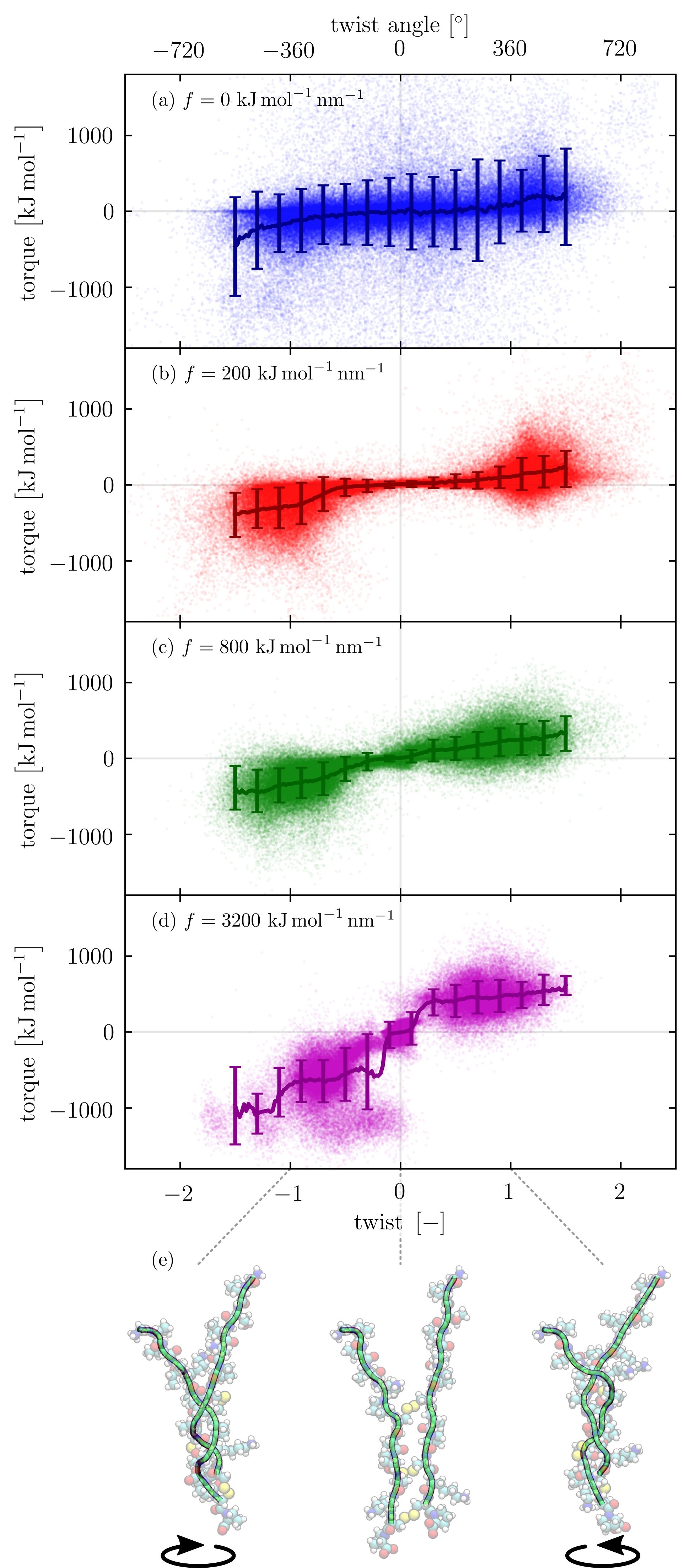}
\caption{The development of the twist-torque profile for increasing stretching force; the exerted torque $\tau$ is shown as a function of the twist $\Tw$ for various stretching forces $f$: (a)~$0$, (b)~$200$, (c)~$800$, and (d) $\SI{3200}{\kilo\joule\per\mole\per\nano\metre}$. (e) Snapshots of the torsion simulation illustrating the conformation of the molecule at specific values of the twist $\Tw=-1,\ 0,\ 1$, for $f = \SI{3200}{\kilo\joule\per\mole\per\nano\metre}$.}
\label{fig:torsion_profiles}
\end{figure}
In Fig.~\ref{fig:torsion_profiles} we show the torsion profiles for various values of the stretching force $f$: (a)~$0$, (b)~$200$, (c)~$800$, and (d)~$\SI{3200}{\kilo\joule\per\mole\per\nano\metre}$. All individual data points are shown with a 5\% opacity, which effectively results in a configuration density plot of the molecule in torque-twist space. The darker lines are trends, which we calculate by averaging the torques $\tau$ for twists $\Tw$ between $-1.5$ and $1.5$, in steps of $\Delta \Tw = 0.02$. The error bars shown are the standard deviations around the averages of the torque. Figure~\ref{fig:torsion_profiles}e shows typical snapshots of the simulations for specific values of the twist $\Tw = -1,\ 0,\ 1$, for $f = \SI{3200}{\kilo\joule\per\mole\per\nano\metre}$.

We learn from Fig.~\ref{fig:torsion_profiles} that regardless of the magnitude of the stretching force $f$, the molecule initially exhibits a torsion stiffening behaviour: the torque $\tau$ increases non-linearly with increasing $\Tw$. For $|\Tw| \lesssim 0.1$ we see that the profile remains fairly flat and close to $\tau \approx 0$, which hints at a low torsional resistance, whereas for increasing values the profile steepens, indicating an increase of the torsional stiffness: the torsional resistance (or ``torsional spring constant'') is associated with the derivative of the torque with respect to the twist. The stiffness increase is in agreement with the results of \citet{VanReenen2013}, who reported torsion stiffening behaviour of the IgG-IgG complex for increasing torsion angles (up to approximately $\SI{250}{\degree}$). In our analysis, however, for even greater rotations the torsion profile appears to flatten off. We return to this phenomenon below. 

We note that for $f=0$, in Fig.~\ref{fig:torsion_profiles}a, the spread of the data points is notably larger than for nonzero stretching forces. This is caused by the fact that the molecule may become strongly supercoiled, and the configuration of the molecule regularly becomes inverted (\textit{i.e.}, the bottom residues end up above the top residues) due to the strong internal stresses caused by the exerted torque and the absence of a stretching force. As a result, the torque needed to reach a certain twist varies rather strongly. Nevertheless, these data points do display characteristics for the mechanical properties of the molecule at zero stretching force. The profile exhibits a mild torsion stiffening and is fairly symmetric for positive and negative values of $\Tw$.

As we increase the stretching force $f$, in Figs.~\ref{fig:torsion_profiles}b and c, the standard deviation decreases and the underlying shape of the torsion profile becomes more apparent: the two chains are pulled towards each other as a result of the stretching and cause the onset of an internal counterforce. The shapes of the two torsion profiles are rather similar, the profile for $f = \SI{800}{\kilo\joule\per\mole\per\nano\metre}$ (Fig.~\ref{fig:torsion_profiles}c) displaying a slightly stiffer profile than the profile for $f = \SI{200}{\kilo\joule\per\mole\per\nano\metre}$ (Fig.~\ref{fig:torsion_profiles}b), as may be expected. For both profiles, we see that in the flat regime, for small twists ($|\Tw| \lesssim 0.1$), the spread is small, compared to the data for stronger twists. This indicates that the torsional stiffness of the molecule is rather insensitive to its exact configuration in the untwisted state. If the molecule is twisted more strongly, its torsional response depends more strongly on how the chains are positioned with respect to each other. While the torsion profiles appear rather symmetric, some features arise that may be associated with the internal structure of the molecule. For example, the magnitude of the torque is greater for negative values of the twist than for positive ones. This hints at a preferred rotation direction for the molecule, which may indicate that the helicity of the structure is slightly right-handed. In addition, we find that ``islands'' form in torque-twist space, around $\Tw = \pm 1$, and that their shape depends on the sign of the twist $\Tw$ and the magnitude of the stretching force $f$.

If we stretch the molecule strongly, with a force $f = \SI{3200}{\kilo\joule\per\mole\per\nano\metre}$, shown in Fig.~\ref{fig:torsion_profiles}d, we find that the features become more pronounced and asymmetric. Still, the profile is rather flat for small values of $\Tw$, however, the aforementioned islands now have a distinctly different shape for positive or negative twists. This indicates that for strong stretching forces $f$, the details in the structure of the molecule may become crucial for the torsional resistance of the protein, and that as a result the asymmetries in the torsion profile arise.

\begin{figure}[htp!]
\centering
\includegraphics[width=\figwidth]{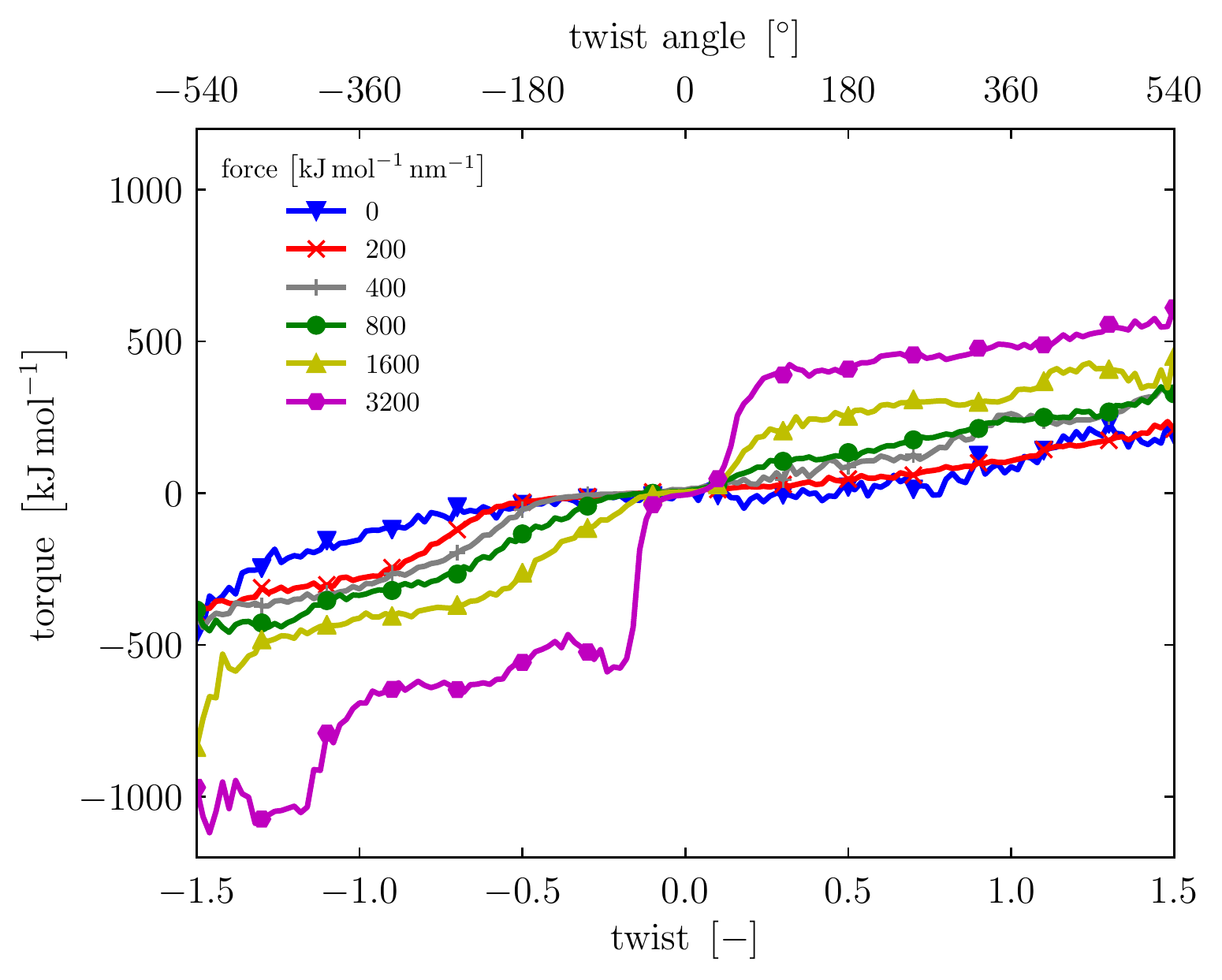}
\caption{The trends of the combined force-torque spectroscopy profile of the IgG linker region: the exerted torque $\tau$ is shown as a function of the twist $\Tw$, for various values of the stretching force $f$. For clarity the individual data points and the standard deviations of the torques are not shown.}
\label{fig:profiles_trendlines}
\end{figure}

If we join together the torsion profiles for different stretching forces $f$, we practically create a combined force-torque spectroscopy analysis of the structure, see Fig.~\ref{fig:profiles_trendlines}.
We only show the trends without the standard deviations, in order to illustrate the overall evolution of the torsional stiffness profile. Note that the full shape of the profile does contain more information about the structural details and the resulting torsional behaviour, and that the shown trends may not necessarily fully represent the shape of the torsion profile, which is apparent from the trendline in Fig.~\ref{fig:torsion_profiles}d. From Fig.~\ref{fig:profiles_trendlines} we learn that the molecule indeed gradually stiffens as the stretching force $f$ increases. It exhibits a torsion stiffening behaviour for small twists, after which the resistance flattens off and the torsional resistance remains approximately constant. Given the double-chain structure of the molecule, we in fact expect the profiles to show an additional torsion stiffening for even greater values of $|\Tw|$ (as the structure becomes more strongly coiled), however we have not investigated greater rotations. We speculate that we do see the onset of such additional stiffening for strongly negative twists in the curves for $f = \SI{1600}{\kilo\joule\per\mole\per\nano\metre}$ and $f = \SI{3200}{\kilo\joule\per\mole\per\nano\metre}$. Figure~\ref{fig:profiles_trendlines} indicates that the end of the torsion stiffening regime, represented by the inflection point of the torque-twist curve, gradually shifts to smaller values of $|\Tw|$ as $f$ increases. In addition, the asymmetry between the magnitude of the exerted torques for positive and negative twists is apparent.

We argue that such an evolution of the shape of the torsion profile of a molecule with increasing stretching force composes a mechanical signature for said molecule. It may facilitate the development of a characterisation method for proteins or other large molecules, for which we can predict the mechanical response by performing this numerical force-torque spectroscopy analysis. 

This concludes the discussion of our results. In the next section, we summarise our main findings and draw our conclusions.

\section*{Conclusion}
We put forward a multiscale molecular simulation method to perform a combined force-torque spectroscopy analysis of large molecules such as proteins: we analyse the mechanical response of a molecule when subjected to external torques and stretching forces. We combine (1) Fluctuating Finite Element Analysis (FFEA) with (2) molecular dynamics (MD) simulations that incorporate external forces. We find that using FFEA, we are able to indicate the region within an IgG molecule that is crucial for its torsional rigidity: while subjecting the molecule to torques, we locate the area of strongest deformation, suggesting that this linker domain is likely to be the most flexible. We subsequently isolate the relevant domain and investigate its torsional properties using MD simulations, while subjecting it to stretching forces and torques. We find that the linker region exhibits a torsion stiffening behaviour, a result that is in qualitative agreement with experimental results by~\citet{VanReenen2013}. For stronger rotations, the exerted torque flattens off and the torsional resistance remains approximately constant. As we increase the stretching force exerted on the molecule, we find that the structure stiffens and that features and asymmetries arise in the shape of the torsion profile, which indicates that the structural details of the molecule may be crucial for its mechanical response. Combining the torsion profiles for different stretching forces effectively results in a combined force-torque spectroscopy analysis of the molecule and we argue that this may be used as a characterisation method for the examined structure.

In conclusion, our study serves as a proof of concept for an efficient numerical evaluation of the mechanical response of a large molecule. This method facilitates the automation of the multiscale procedure for a high-throughput computational analysis of multiple proteins subject to stretching and torsional forces. If the atomistic structure of such a molecule is known, we may disregard the domains less relevant to the torsional stiffness and focus on the softest areas from the perspective of the torsional rigidity, aided by the FFEA method. Bulky domains within the molecule that consist of many atoms and are likely to be relatively rigid, do not need to be taken into account. This is a considerable advantage for the atomistic molecular dynamics simulation that is subsequently used to investigate the torsional properties of the relevant domain in more detail.

\section*{Author contributions}
TWGvdH performed simulations, analysed the results and wrote the manuscript. DJR, OGH and SAH developed the FFEA simulation method. TWGvdH, PvdS, SAH and CS designed the research and interpreted the results. All authors contributed to the manuscript.

\section*{Acknowledgments}
Molecular images were produced using the VMD software package~\cite{Humphrey1996}. The authors thank Ben Hanson, Fabiola Guti\'errez-Mej\'ia and Ren\'e de Bruijn for fruitful discussions. TWGvdH thanks the University of Leeds, where part of this work was executed, for their hospitality.

\section*{Appendix}
\subsection*{The elastic stress}
\label{app:elastic_stress}
The elastic energy of the molecule is based on a Mooney-Rivlin hyperelastic model for the stress~\cite{Shontz2012}, with an adaptation proposed by~\citet{Gent1996}, in order to introduce a maximum deformation for the structure. The strain energy density function $W$ is as follows:
\begin{equation}
\label{eq:MR_strain_energy}
\begin{split}
W = &-\frac{G}{2}(\Im - 3)\ln \left(1 - \frac{I_1-3}{\Im-3}\right) \\
&+\frac{3K-2G}{12}\left(J^2-1\right)-\frac{3K+4G}{6}\ln J,
\end{split}
\end{equation}
with $G$ and $K$ the shear and bulk moduli, respectively, and $I_1$ and $J$ two invariants of the deformation gradient tensor $\pmb F$:
\begin{align}
I_1&=\mathrm{Tr}\left(\pmb F\pmb F^T\right);\\
J &= \mathrm{Det}\left(\pmb F\right) = {V}/{V_0},
\end{align}
with $V$ and $V_0$ the instantaneous and initial volumes, respectively. $\Im$ is the maximum value for $I_1$ for the structure. In our analysis, in order to strongly limit the deformations and to emphasise the stressed regions, we set $\Im=3.1$. $\pmb F$ describes the deformation for a mapping of a position $\pmb X$ to a new position $\pmb x$: $\pmb X \mapsto \pmb x(\pmb X)$:
\begin{equation}
F_{ij}=\pdd{x_i}{X_j}.
\end{equation}
The elastic stress tensor is derived from the strain energy density function as follows:
\begin{equation}
\pmb{\sigma}_\mathrm{e} = \frac{1}{J}\pdd{W}{\pmb F}\pmb F^T,
\end{equation}
which results in:
\begin{equation}
\pmb{\sigma}_\mathrm{e} = \frac{G}{J} \frac{\Im - 3}{\Im - I_1} \pmb F\pmb F^T+\left(\frac{3K-2G}{6J}\left(J^2-1\right)-\frac{G}{J}\right)\pmb I,
\end{equation}
with $\pmb I$ the identity matrix.

\subsection*{Calculating the twist}
\label{app:calc_twist}
\begin{figure}[htp!]
\centering
\def\svgwidth{\figwidth}
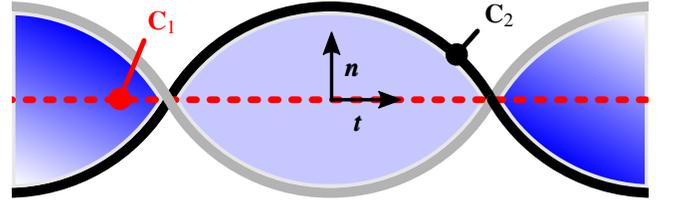
\caption{A ribbon with central axis curve $\mathbf{C}_1$ and boundary curve $\mathbf{C}_2$. The unit tangent to the central axis $\pmb{t}$ and the unit normal vector $\pmb{n}$ pointing from $\mathbf{C}_1$ to $\mathbf{C}_2$ are indicated. This particular ribbon contains a twist $\Tw = 1$.}
\label{fig:ribbon_twist}
\end{figure}

The twist of a ribbon indicates the amount of (right-handed) winding of the ribbon around itself, along the central axis. We can define a ribbon by considering the central axis $\mathbf{C}_1$ and the boundary $\mathbf{C}_2$, see Fig.~\ref{fig:ribbon_twist}.

\begin{figure}[htp!]
\centering
\def\svgwidth{\figwidth}
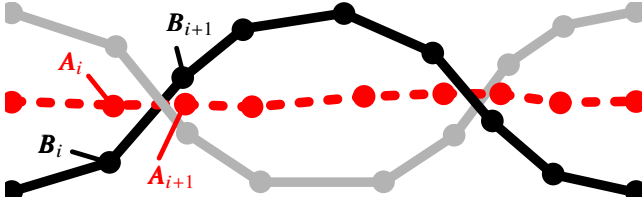
\caption{A discrete ribbon with central axis positions $\pmb{A}_i$ and boundary positions $\pmb{B}_i$.}
\label{fig:ribbon_disc}
\end{figure}

We indicate the unit tangent to the central axis $\pmb{t}(s)$ and the normal unit vector $\pmb{n}(s)$ perpendicular to $\pmb{t}$, pointing from $\mathbf{C}_1$ to $\mathbf{C}_2$. They share a common arc length parameter $s$. The twist $\Tw$ of the ribbon can be calculated as~\cite{Au2008},
\begin{equation}
\label{eq:twist}
\Tw = \frac{1}{2 \pi} \int_{\mathbf{C}_1} \left(\pmb{t} \times \pmb{n}\right)\cdot \dd{\pmb{n}}{s} \d{s}.
\end{equation}
In order to calculate the twist contained in the linker region (see Fig.~\ref{fig:IgG_linker}), we first need to transform this region to a ribbon representation. We consider only the backbones (consisting of carbon and nitrogen atoms) from the lowest disulfide bond up to the top. These are the atoms corresponding to the green ribbons in Fig.~\ref{fig:torsion_profiles}e. Taking into account the dangling ends at the bottom of the two chains in the twist calculation would result in overestimations of the intrinsic twist of the protein, since these are hindered less by the two-chain structure. As the two chains are structurally identical, it is prudent to connect the corresponding atoms in order to form a ribbon. We define the central axis $\pmb{A}_i$ as the average positions of the connected atoms in each chain,
\begin{equation}
\pmb{A}_i = \frac{\pmb{r}_{1,i} + \pmb{r}_{2,i}}{2},
\end{equation}
where $\pmb{r}_{j,i}$ denotes the position of the $i$-th atom in chain $j$. The discrete ribbon is now defined by the central axis $\pmb{A}_i$ and the boundary $\pmb{B}_i \equiv \pmb{r}_{1,i}$, see Fig.~\ref{fig:ribbon_disc}.

We calculate the twist $\Tw$ of the discrete ribbon as
\begin{equation}
\label{eq:twist_disc}
\Tw = \frac{1}{2 \pi} \sum_{i=1}^{n-1} \alpha_i \arccos\left( \v_i \cdot \w_i \right),
\end{equation}
where
\begin{align}
\alpha_i &= \sgn\left[\A_i \A_{i+1} \cdot \left(\v_i \times \w_i \right)\right]; \\
\v_i &= \frac{\B_i \A_i \times \B_i \A_{i+1}}{||\B_i \A_i \times \B_i \A_{i+1}||}; \\
\w_i &= \frac{\B_{i+1} \A_i \times \B_{i+1} \A_{i+1}}{||\B_{i+1} \A_i \times \B_{i+1} \A_{i+1}||},
\end{align}
and $\A_i \A_{i+1}$ indicating the segment from position $\A_i$ to $\A_{i+1}$~\cite{Au2008}. The factor $\alpha_i$ accounts for the direction of the twist, \textit{i.e.}, whether the respective segments cause a positive or negative contribution to the total twist.

\subsubsection*{Considering the writhe}
In our analysis of the rotation of the molecule we also considered using the writhe of the structure. In order to calculate the writhe, we construct a closed loop consisting of the two backbones of the chains, the lowest disulfide bond connecting the two chains and an artificial connection between the top two residues. The writhe is a quantity of a closed loop, which means that it is independent of any external reference axis or orientation. However, it turns out that the writhe is rather sensitive to fluctuations in the positions of the atoms. This results in uncertainties and inaccuracies in determining the rotation of the molecule and we therefore deem it unsuitable for our analysis.

\bibliography{literature}

\begin{thebibliography}{40}
\bibitem[Not()]{Note}
Using FFEA to perform a full analysis of the torsional rigidity of the molecule
  turns out to not be possible: for high values of $\phi$ (resulting from high
  values of the torque $\tau$), the molecule relaxes its internal stresses by
  twisting back the linker region and thereby moving through its own surface,
  which is a non-physical phenomenon. Unfortunately, incorporating a steric
  surface-surface repulsion for the molecule in order to prevent this results
  in computationally infeasible calculations at late times: the linker region
  coils up and much of the surface remains at a close distance to itself. The
  simulations eventually fail because the molecule cannot relax its internal
  stresses and as a result the elements in the linker invert due to the strong
  forces on the nodes.
\end{thebibliography}

\end{document}